\documentclass[12pt]{article}
\usepackage{epsfig}
\def\be{\begin{equation}}
\def\ee{\end{equation}}
\def\ben{\begin{displaymath}}
\def\een{\end{displaymath}}
\def\ba{\begin{array}{c}}
\def\bal{\begin{array}{l}}
\def\ea{\end{array}}
\def\p{\partial}

\begin{document}


\vspace{1.5cm}

\begin{center}
{\Large {\bf

Conditional observability versus self-duality
 in a schematic model

 }}
\end{center}

\vspace{10mm}

\begin{center}
{\bf Miloslav Znojil}

\vspace{3mm} \'{U}stav jadern\'e fyziky AV \v{C}R,

250 68 \v{R}e\v{z},

Czech Republic

{e-mail: znojil@ujf.cas.cz}

\vspace{3mm}

\vspace{5mm}

\end{center}

\vspace{5mm}


\section*{Abstract}

For a 4-dimensional 3-parametric toy Hamiltonian $H(a,b,c)$ we
construct the domain ${\cal D}$  of couplings in which the
eigenvalues $E_n$ remain real (i.e., in principle, observable). A
relationship is found between the reflection symmetry of the
spectrum (i.e., its Dunne's and Shifman's self-duality $
E_j=-E_{N+1-j}$ at $N=4$) and a geometric symmetry of the physical
domain ${\cal D}$. Simultaneously, {\em both} remain unbroken at
$a=c$.

\newpage

\section{Introduction}

Reflection symmetry
 \be
 E_j=-E_{N+1-j}\,,\ \ \ \ j = 1, 2, \ldots \,
 \label{selfdual}
 \ee
between the low- and high-excitation parts of the energy spectra
is one of features of all the many-body Hamiltonians which are
quadratic in the creation and annihilation operators
\cite{Blaizot}. In a different physical context of quasi-exactly
solvable local potentials, Dunne and Shifman~\cite{DS} found the
same form of symmetry and called it self-duality of the spectrum.
Recently, it was rather surprising for us to detect the emergence
of the same spectral symmetry (\ref{selfdual}) also during the
study of the strong-coupling version of a family of
finite-dimensional phenomenological non-Hermitian (often called
${\cal PT}-$symmetric) chain-model Hamiltonians \cite{maximal}.

The latter study revealed that the additional symmetry
(\ref{selfdual}) appeared in connection with a thorough
simplification of the spectra {\em and also} with a facilitated
feasibility of an algebraic determination of the positions of
certain maximal physical couplings. Formally, the latter maximal
couplings were defined as vertices of the boundary $\p {\cal
D}^{}$ of the domain ${\cal D}^{}$ where the free parameters of
our models remained physically acceptable, i.e., compatible with
the reality and observability of the energies. As long as the
domain ${\cal D}$ proved bounded, the observability of the related
energies can be called ``conditional" \cite{condit}. In our
subsequent deeper study~\cite{II} of the same class of models with
symmetry (\ref{selfdual}) we succeeded in a further constructive
extension of the previous result and in a clarification of the
geometric structure of the horizons $\p {\cal D}^{}$ in non-empty
vicinities of all the above-mentioned vertices. In our present
brief communication we feel inspired by the latter experience.

Our purpose is to throw new light on the peculiar apparent
correspondence between the spectral symmetry (\ref{selfdual}) and
the geometric shape of the boundary $\p {\cal D}$. An explicit
description of the latter manifold will be performed here for a
specific, modified model where the up-down symmetry
(\ref{selfdual}) will be manifestly broken. We shall see, in
essence, that after the violation of symmetry (\ref{selfdual})
also the mysterious simplicity of the shape of the surface $\p
{\cal D}^{}$ gets lost. For the sake of clarity of our argument,
we shall only consider the construction of $\p {\cal D}$ for the
exactly solvable three-parametric matrix Hamiltonian
 \begin{equation}
H=\left (
\begin{array}{cccc}
-3 & b & 0 & 0 \\ -b & -1 & a & 0 \\ 0 & -a & 1 & c \\ 0 & 0 & -c
& 3
\end{array}
\right )\,.  \label{nemarnit}
 \end{equation}
The problem differs from its two-parametric $b=c$ predecessor of
ref.~\cite{maximal} (cf. also a few related marginal remarks in
\cite{determ}) by a non-vanishing measure $f>0$ of the extent of
the asymmetry which enters the upper coupling defined as
$b=b(c,f)=\sqrt{c^2+f^2}$.

We shall not analyze the possibilities of a transition to the
higher-dimensional models here. A thorough discussion of such a
perspective has already been presented in both our preceding
papers on self-dual cases~\cite{maximal,II}. Basically, we saw
there that the key information about the structure of the spectra
is already provided by the models of the lowest dimensions. The
growth of the matrix dimension appeared to provide merely
technical alterations of the overall picture.

On this background, we now intend to pay more attention to the
most elementary up-down-asymmetric model (\ref{nemarnit}). We
shall apply computer-assisted linear algebra and formulate some of
its consequences in sections \ref{analysis} and \ref{alysis}, with
a brief summary in section \ref{summary}.


\section{Physical domain ${\cal D}$
(graphical approach) \label{analysis}}

Let us start by a terminological remark concerning eq.
(\ref{nemarnit}) and all its generalizations with parameters
inside ${\cal D}$. Its reason is that the accepted denotation of
similar models characterized by a non-standard transposition
asymmetry differs in different applications. We will prefer
calling our $H$ quasi-Hermitian (a term typical for nuclear
physics \cite{Geyer}). Still, several authors also use the
equivalent names of pseudo-Hermitian $H$ (predominantly in the
context of linear algebra and mathematics \cite{Ali}) or
crypto-Hermitian $H$ (this nice and most self-explanatory concept
appeared recently in the context of gauge models \cite{Smilga}).
Last but not least, another, less strict nickname of ${\cal
PT}-$symmetric $H$ as coined by Carl Bender~\cite{Carl} became
most popular as particularly appealing in field theory.

%


\subsection{The allowed range of $A(a,c,f)$.}

Returning now to our particular model let us recollect that its
energies coincide with the roots of its secular determinant,
\begin{equation}
\det \left (H-E\,I \right ) =0\,,  \label{prob4f}
\end{equation}
where we decided to set $b^2=c^2+f^2$, with a new measure of the
up-down asymmetry $f>0$ replacing the original coupling parameter
$b$. We may express eq.~(\ref{prob4f}) in its explicit polynomial
form
  \be
{E}^{4}_{1,2,3,4}-A(a,c,f)\,
 {E}^{2}_{1,2,3,4}
+C(a,c,f)=4\,f^2\, {E}_{1,2,3,4}\,
 \label{secu4}
  \ee
where we have $\ A(a,c,f)= 10-{a}^{2}-2\,{c}^{2}-{f}^{2}\ $ while
  \ben
 C(a,c,f)=9+6\,{{\it c}}^{2 }-9\,{{\it
 a}}^{2}+3\,{{\it f}}^{2}+{{\it c}}^{4}+{{\it f}}^{2}{{ \it
 c}}^{2}
 \een
or, in compact form,
 \ben
  C(a,c,f)=
 \left (
 3+c^2\right )^2
 +f^2\,
 \left (
 3+c^2\right )
 -
 9\,a^2=
 \left (
 3+c^2+f^2/2\right )^2
 -
 \left (
 9\,a^2+f^4/4
 \right)=
 \een
 \ben
 =
 \left (
 8-A/2-a^2/2 \right )^2
 -
 \left (
 9\,a^2+f^4/4 \right)\,.
 \een
\begin{figure}[h]                     
\begin{center}                         
\epsfig{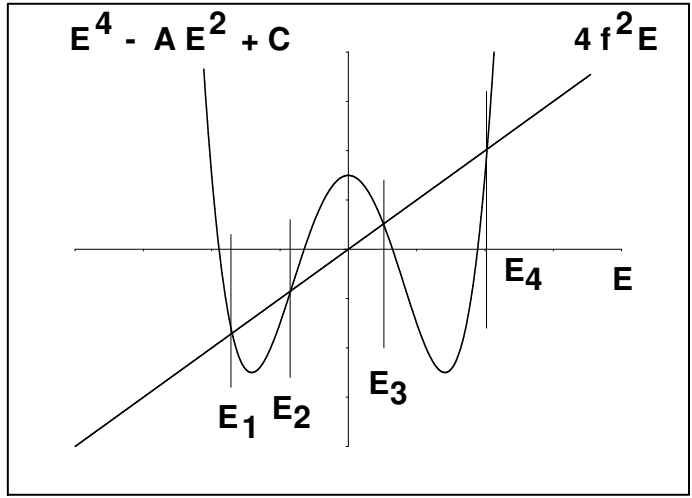}
\end{center}                         
\vspace{-2mm} \caption{Graphical solution of our secular
eq.~(\ref{secu4} at $A>0$).
 \label{obr1a}}
\end{figure}
We observe that at any positive $f>0$ and negative (or vanishing)
$A\leq 0$, the left-hand-side curve of eq.~(\ref{secu4}) can only
have {\em at most two} real intersections $E_{1,2}=E_{1,2}(f)$
with the right-hand-side straight line. This means that the
spectrum {\em always} contains at least two complex conjugate
energy roots and we stay off ${\cal D}$.

If we allow $A\leq 0$ and vanishing $f=0$, the two complex roots
become real only in the limit $A \to 0$. Thus, the point of the
boundary $\p {\cal D}$ of the physical domain of parameters would
be achieved, in full agreement with our older self-dual results
\cite{maximal}. Now, we decided to assume that the self-duality is
manifestly broken,  $f>0$, so that we may conclude that we have
{\em always} to consider just the strictly positive values of the
polynomial
 \be
 A(a,c,f) > 0\,,
  \ \ \ \ \ \  (a,c,f) \in {\cal D}  \,.
 \ee
Under this assumption, the graphical solution of eq.~(\ref{secu4})
is sampled, in arbitrary units, in Figure \ref{obr1a}. At a
variable $f>0$ we shall further require that inside ${\cal D}$,
the left-hand side superposition of the two (viz., quadratic and
quartic) parabolic curves of eq.~(\ref{secu4}) has strictly four
real intersections with the right-hand side straight line. From
this assumption one immediately deduces that we must have a
negative $E_1(f)>E_1(0)$ and $E_2(f)<E_2(0)$ and a positive
$E_3(f)<E_3(0)$ and $E_4(f)>E_4(0)$. Hence, the two rightmost
roots $E_3$ and $E_4$ can never merge and, subsequently, they
cannot form a complex conjugate doublet unless $f=A=0$. In terms
of the geometry of ${\cal D}$, the transition to $f \neq 0$
immediately destroys the multidimensional hedge-hog-shaped form of
the surface $\p {\cal D}$ as observed at $f=0$ \cite{II}.

\subsection{The allowed range of $C(a,c,f)$.}

An inspection of Figure \ref{obr1a} reveals that at $E=0$, one of
the curves acquires the value of $C(a,c,f)$ which may be, in
principle, positive, vanishing or negative. Once we keep the
quantity $f=f_0>0$ fixed, the variation of $a$ and/or $c$ just
changes the values of the functions $C(a,c,f_0)$ and $A(a,c,f_0)$.
In the two-parametric space of $a$ and $c$ we may move along a
line of a constant $A(a,c,f_0)$ and observe that any decrease of
$C(a,c,f_0)$ leads to the decrease of the distance between the
intersections $E_2(f_0)$ and $E_3(f_0)$. In principle, these two
values will merge and subsequently complexify below certain
$f_0-$dependent value $ C_{(minus)}(a,c,f_0)<0$. Similarly, the
growth of the value of the function $C(a,c,f_0)$ will imply the
decrease of the distance between the two leftmost energies
$E_1(f_0)$ and $E_2(f_0)$. The latter pair will merge (and
subsequently complexify) at an upper bound $ C_{(plus)}(a,c,f_0)$
which, by the way, need not be necessarily positive. We may
conclude that all the four energies remain real iff
 \be
 C_{(minus)}(a,c,f) \leq C(a,c,f) \leq C_{(plus)}(a,c,f)
 \,, \ \ \ \ \ \  (a,c,f) \in {\cal D}\,.
 \label{inequ}
 \ee
Both these bounds follow also algebraically from our secular
equation (\ref{secu4}) of course. Their occurrence corresponds to
the confluence of the above-mentioned pairs of energies at a point
of the boundary $\partial {\cal D}$ of ${\cal D}$.

In the graphical language this means that inside all the interior
of ${\cal D}$, the graph of the secular polynomial $Y(E)=\det
(H-E)= E^4-A\,E^2-4f^2\,E+C$ will intersect the real line four
times, achieving its leftmost, negative minimum at a negative real
variable $z_{(min)}\in (E_1,E_2)$,
 \be
 Y(z_{(min)})=z_{(min)}^4-A\,z_{(min)}^2
 -4f^2\,z_{(min)}+C_{(plus)}<0\,.
 \ee
The subsequent positive maximum of $Y(z)$ will occur at a larger
$z_{(max)}\in (E_2,E_3)$,
 \be
 Y(z_{(max)})=z_{(max)}^4-A\,z_{(max)}^2
 -4f^2\,z_{(max)}+C_{(minus)}>0\,.
 \ee
As long as we showed above that the two rightmost energy roots
$E_3$ and $E_4$ can never merge, the remaining, third extreme of
the function $Y(z)$ at some $z\geq z_{max}$ may and will be
ignored here as redundant.
\begin{figure}[h]                     
\begin{center}                         
\epsfig{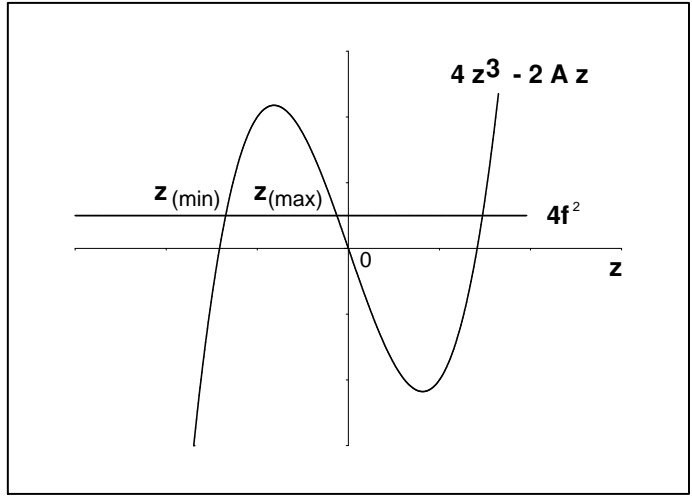}
\end{center}                         
\vspace{-2mm} \caption{Graphical solution of eq.~(\ref{sesscu4})
at $A>0$.
 \label{obr2}}
\end{figure}

With the variation of $a$, $c$ and $f$, we shall always have
$Y(0)\,\equiv\,C$ while the quantities $z_{(min)} \leq z_{(max)}$
correspond to the respective leftmost minimum and subsequent
maximum of the difference curve $Y(z)$ (in Figure \ref{obr1a}, you
may imagine that $f$ varies). Algebraically, the two quantities
$z_{(min)}$ and $z_{(max)}$ coincide with the two real roots of
the third-order polynomial $\p Y/\p_z$, i.e., of the cubic
equation
 \be
4\,{z}^{3}_{\tiny
  \left (\!\!\!\!
  {\begin{array}{l} max \\ \noalign{\vspace{-1mm}} min \ea}
  \!\!\!\!
  \right )} -2\,A(a,c,f)\,
 z_{\tiny
  \left (\!\!\!\!
  {\begin{array}{l} max \\ \noalign{\vspace{-1mm}} min \ea}
  \!\!\!\!
  \right )}
 =4\,f^2\,
 \label{sesscu4}
  \ee
(cf. Figure \ref{obr2}). Once we know these auxiliary values, we
can define the required bounds from their definition
 \be
 C_{\tiny  \left (\!\!\!\!
 {\begin{array}{l} minus \\
 \noalign{\vspace{-1mm}}  plus \ea}\!\!\!\!  \right )
  }=4\,f^2\, {z}_{\tiny  \left (\!\!\!\!
 {\begin{array}{l} max \\ \noalign{\vspace{-1mm}}  min \ea}
 \!\!\!\!
  \right )}
 + A\,
 {z}^{2}_{\tiny
  \left (\!\!\!\!
  {\begin{array}{l} max \\ \noalign{\vspace{-1mm}} min \ea}
  \!\!\!\!
  \right )}-
 {z}^{4}_{\tiny
 \left (\!\!\!\!  {\begin{array}{l} max \\ \noalign{\vspace{-1mm}}
 min \ea}\!\!\!\!  \right )}
  \,
 \label{secddu4}
  \ee
which, after an algebraic simplification using
eq.~(\ref{sesscu4}), degenerates to the simpler formula
 \be
 C_{\tiny \left (\!\!\!\! {\begin{array}{l} minus \\
  \noalign{\vspace{-1mm}}
   plus \ea}\!\!\!\!  \right )
 }=\frac{A(a,c,f)}{2}\,{z}^{2}_{\tiny
  \left (\!\!\!\!  {\begin{array}{c} max \\
  \noalign{\vspace{-1mm}}   min \ea}\!\!\!\!  \right )
 }
 +3\,f^2\,
 {z}^{}_{\tiny  \left (\!\!\!\!
 {\begin{array}{c} max \\  \noalign{\vspace{-1mm}}  min \ea}
 \!\!\!\!
  \right )
 }\,.
 \label{resulting}
 \ee
We may conclude that since the only negative component of the
polynomial $C(a,c,f)$ is $-9\,a^2$, the latter bounds in effect
fix the allowed range of the parameter~$a$ for any given pair of
$c$ and $f$. In the other words, the middle energy levels $E_2$
and $E_3$ can merge iff the value of $a$ proves sufficiently
large. Moreover, since $A>0$, the quadruple merger of $E_1$ and
$E_2$ and $E_3$ with $E_4$ would require a return to $f=0$ and
proves entirely excluded at $f \neq 0$.

\section{Physical domain ${\cal D}$
(analytic approach) \label{alysis}}

In our previous studies \cite{maximal,II} of $N-$dimensional chain
models it has been conjectured and verified that at all $N$, the
postulate of a self-duality of the spectrum $ E_j=-E_{N+1-j}$
could perceivably simplify the construction as well as geometric
characterization of the shape of the manifold ${\cal D}$. In the
preceding section \ref{analysis} an independent constructive
support of the plausibility of such a relationship has been given
via our three-parameter non-selfdual example (\ref{nemarnit}). In
what follows we intend to complement these results by paying
attention to a simplification and non-numerical description of the
boundary $\p {\cal D}$. This might prove relevant, e.g., in the
quantitative context of the strong-coupling and/or singular forms
of perturbation theory \cite{II} as well as in a more qualitative
setting of a possible quantum version of theory of
catastrophes~\cite{condit}.

\subsection{The allowed range of the asymmetry $f$.}

We assumed the knowledge of the input
energy-degeneracy-determining auxiliary quantities $z_{(min)}$ and
$ z_{(max)}$ which can both be defined by the Cardano formulae in
principle. Fortunately, a perceivably less painful recipe can be
recommended. In a preparatory step, an inspection of
eq.~(\ref{sesscu4}) (cf. also Figure \ref{obr2}) reveals that the
pair of our auxiliary quantities $z_{(min)} \leq z_{(max)}$
remains real iff $f \in (0,f_{(upper)})$. Here, the numerical
value of the upper bound $f_{(upper)}$ annihilates the polynomial
$Z(z)=4\,z^3-2\,A\,z-4\,f^2$ at its left maximum where $\p Z/\p_z
=0$ at $z=z_{(upper)}$. This means that
 \ben
 12\, z_{(upper)}^2=2\,A\,,\ \ \ \ \ \
 z_{(upper)}=-\sqrt{\frac{A}{6}}\,,\ \ \ \ \ \
 f_{(upper)}=\left ( \frac{A^3}{54}\right )^{1/4}\,.
 \een
Thus, we are allowed to re-parametrize $f \to \varphi$ in such a
way that
 \ben
 f^2=f_{(upper)}^2\,\cos \varphi\,,\ \ \ \ \ \
 \varphi =\varphi(A,f)\,\in (0, \pi/2)\,.
 \een
Moreover, using the direct insertion in eq.~(\ref{sesscu4}) we
easily verify that its closed solutions simply read
 \ben
 z_{(min)}
 =-\sqrt{\frac{2A}{3}}\,
 \cos \left (\frac{\pi-\varphi}{3}
 \right )\,,
 \een
 \ben
 z_{(max)}
 =-\sqrt{\frac{2A}{3}}\,
 \cos \left (\frac{\varphi+\pi}{3}
 \right )\,.
 \een
Both  of them are negative and they only have to satisfy the pair
of constraints
 \be
 z_{(min)}
 =z_{(min)}(A,\varphi)
   \in (-\sqrt{A/2}, -\sqrt{A/6})\,,
 \ee
 \be
 z_{(max)}
 =z_{(max)} (A,\varphi)
 \in (-\sqrt{A/6},0)\,.
 \ee
This enables us to conclude that all the above-mentioned open
interval of $\varphi$ may be treated as lying inside ${\cal D}$.
The angle $\varphi$ is, therefore, a better measure of the
breakdown of the self-duality of our toy Hamiltonian $H$. Next,
let us show that and why also its further coupling constants
should be reparametrized.

\subsection{A reparametrization
$(a,c,f) \to (\alpha,\delta,\varphi)$ of $H$ and ${\cal D}$}

Let us replace $A \to \alpha$ in such a way that $A=
10\,\sin^2\alpha$ with $\alpha \in (0,\pi/2)$. Moreover, let us
also reparametrize $a\to \delta$ with $a^2= 10\,\cos^2\alpha\,
\sin^2\delta$ and $\delta \in (0,\pi/2)$. This implies that the
resulting formula $2\,c^2+f^2= 10\,\cos^2\alpha\, \cos^2\delta$
may be inserted in the definition of
 \ben
 C(a,c,f)=
 \left (
 3+5\,\cos^2\alpha\, \cos^2\delta\right )^2
 -
 90\,\cos^2\alpha\,
 \sin^2\delta-f^4/4=
 \een
 \ben
 = \left (
 12+5\,\cos^2\alpha\, \cos^2\delta\right )^2
 -90\,\cos^2\alpha -135-f^4/4\,.
 \een
In parallel, we may also reparametrize eq.~(\ref{resulting}),
 \be
 C_{\tiny \left (\!\!\!\! {\begin{array}{l} minus \\
  \noalign{\vspace{-1mm}}
   plus \ea}\!\!\!\!  \right )
 }=
 \frac{100}{3}\,\sin^4\alpha\,
  \cos \left (\frac{\pi \pm \varphi}{3}
 \right ) \left [
  \cos \left (\frac{\pi \pm \varphi}{3}
 \right )
 - \cos \varphi\,\right ]
 \,
 \label{presult}
 \ee
which is independent of $\delta$. Thus, we may introduce another
pair of shifted and safely positive boundaries
 \ben
 B_{\tiny \left (\!\!\!\! {\begin{array}{l} minus \\
  \noalign{\vspace{-1mm}}
   plus \ea}\!\!\!\!  \right )
 }=
 C_{\tiny \left (\!\!\!\! {\begin{array}{l} minus \\
  \noalign{\vspace{-1mm}}
   plus \ea}\!\!\!\!  \right )
 }+
 90\,\cos^2\alpha +135+f^4/4 \ \equiv\
 B_{\tiny \left (\!\!\!\! {\begin{array}{l} minus \\
  \noalign{\vspace{-1mm}}
   plus \ea}\!\!\!\!  \right )
 }(\alpha,\varphi)\,
 \een
and arrive at the final and amazingly compact form of our key
constraint (\ref{inequ}),
 \be
 \sqrt{B_{(minus)}(\alpha,\varphi)} \leq
  12+5\,\cos^2\alpha\, \cos^2\delta
  \leq \sqrt{B_{(plus)}(\alpha,\varphi)}
 \,, \ \ \ \ \ \  \delta \in {\cal D}\,.
 \label{finineq}
 \ee
Our construction of the quasi-Hermiticity \cite{Geyer} domain
${\cal D}$ is completed since eq.~(\ref{finineq}) is its
definition. In the limit $f\to 0$, this equation also has been
checked to specify the simpler domain ${\cal D}$ described in
paper \cite{maximal}.

At $f > 0$ our present generalized definition~(\ref{finineq}) of
${\cal D}$ is algebraic and rigorous but its geometric
interpretation is not too transparent. Indeed, although our two
inequalities specify the allowed, physical range of $\delta$ for
any given pair of parameters $(\alpha,\varphi)$, the interval can
prove empty. This would mean that the domain $ {\cal D}$ does not
contain any point with pre-selected $(\alpha,\varphi)$.
Unfortunately, an account of similar subtleties already lies a bit
beyond the scope of our present brief communication.

%

\section{Summary \label{summary}}

Any non-Hermitian ${\cal PT}-$symmetric quantum Hamiltonian $H$
remains physical in a domain of parameters ${\cal D}$ where the
spectrum is real and, hence, measurable. In refs.
\cite{maximal,II} we revealed that several properties of a family
of tridiagonal matrix Hamiltonian models of this type become
exceptionally transparent after an imposition of the up-down
symmetry or ``self-duality" requirement (\ref{selfdual}). In our
present study we complemented this observation by a test of the
consequences of a manifest violation of the self-duality.

We saw that once the real measure $f^2$ of the violation of
self-duality becomes different from zero, our first nontrivial,
up-down-asymmetric model (\ref{nemarnit}) (skipped in
\cite{determ} as complicated) offers a challenging eigenvalue
problem. We have shown that its discussion and the non-numerical
construction of its physical domain of parameters ${\cal D}$ still
remains feasible and compact.

We believe that our present construction of a toy model with
non-numerically tractable physical horizon $\p {\cal D}$ will
further improve our understanding of the mathematics which
underlies the so called ``conditional-solvability" phenomenon
\cite{condit} as well as the analyses of the so called ``quantum
catastrophes" \cite{condit,II,conditbe}. At present, the {\em
practical} role of these fairly fresh mathematical concepts finds
new and new physical applications, the number of which is rapidly
increasing. Besides their above-stressed innovative role in
quantum theory and particle physics \cite{Carl}, it is worth
noting, in the conclusion, that their use also currently inspired
new progress in the areas as remote as relativistic cosmology
\cite{Ali} and phenomenological magnetohydrodynamics \cite{Uwe}.

\section*{Acknowledgement}

Work supported by GA\v{C}R, grant Nr. 202/07/1307.



\section*{Figure captions}

\subsection*{Figure 1. Graphical solution of our secular
eq.~(\ref{secu4}) at $A>0$.}

\subsection*{Figure 2. Graphical solution of eq.~(\ref{sesscu4}) at $A>0$.}

\end{document}